# Layered compounds AFBiS$_2$: superior birefringent crystals

Hai Wang[*]

College of Materials Science and Engineering, Tongji University, Shanghai 201804, China

Abstract

Stimulated by the discovery of the giant birefringence (GBF) in LaOBiS$_2$ (H. Wang, arxiv: 1304.5032), we check SrFBiS$_2$ and find a superior birefringence up to 1.28. Naturally, AF layer (i.e., SrF) is considered to be of important and is demonstrated by the further study on AFBiS$_2$ (A=Mg, Ca, Ba). Interestingly, MgFBiS$_2$ is found to have the largest birefringence of about 1.60. Furthermore, the origin of GBF in AFBiS$_2$ is discussed in detail based on first-principles calculations on the electronic structure and refractive index.

[*] Corresponding author, E-mail addresses: hwang@tongji.edu.cn

Birefringence (BF) is the optical anisotropy in refractive index due to the different light propagation velocity within the material. This effect was first found in calcite by Danish scientist Rasmus Bartholin in 1669.[1] There are three well-known natural birefringent materials: calcite $CaCO_3$ ( n=0.172), quartz $SiO_2$ ( n=0.009), and $MgF_2$ ( n=0.006), all measured at 590 nm wavelength.[1] However, the birefringence is so small that its application is limited severely. On the other hand, birefringence has been extensively used in imaging spectrometer, laser devices,[2] optical components, photonic crystal fibers[3] and quantum metrology[4]. Hence, seeking new birefringent materials is of important for both fundamental research and industrial application.[5-6,7] There are at least two popular approaches: one is artificial crystal and the other is nanostructure engineering.

Presently, nanostructure engineering seems to be only one effective approach. For example, the in-plane birefringence of nanostructured ZnO nanowire arrays is about one order of magnitude higher than that of quartz.[8] And then, nanostructured GaP nanowires are found to have a giant birefringence (0.8),[9] which is enhanced up to 75 times than that of quartz. The giant birefringence (~ 0.2) is also report in anisotropically nanostructured silicon.[10] Through dynamic control of growth geometry, the birefringence of nanostructured Si thin-films has a maximum up to 0.4, which is comparable to two-dimensional photonic crystals.[11] Naturally, Glancing Angle Deposition (GLAD)[12] is developed to be and effective method for growth excellent birefringent nanostructured materials.

On the other hand, artificial crystal has a relative small birefringence. For example, $YVO_4$ has a birefringence about 0.22, which is inferior to rutile $TiO_2$ (0.28). But $YVO_4$ has been extensively used in laser devices and optical components owing to its easy growth, large refractive index and good mechanical properties.[13] Recently, $LaOBiS_2$ is predict to have a giant birefringence of 1.0, which is about 111 times larger than that of quartz.[14] This is interesting because the single-crystal compound is first-time shown to have a superior BF than that of nanostructured materials through nanostructure-engineering. Whatever, the origin of GBF is unclear so far. In this work, we check the birefringence of $AFBiS_2$ and discussed its origin in detail.

The calculations in this work were performed using density functional theory (DFT). The more details can be found in Ref. [14]. We first carried out the structural calculations on $AFBiS_2$ and listed the results in Table I. For $SrFBiS_2$, its lattice constant a is 1.1% larger than the experimental value, while the lattice constant c is 2.2% smaller. Similar situation has been report for $LaOBiS_2$.[15] Our results of $LaOBiS_2$ agree well with previous experimental report.[16] Both lattice constants a and c of $AFBiS_2$ from Mg to Ba are increase as A-ions radii increasing. This is expected because the larger ionic radii the larger lattice constant. On the other hand, $AFBiS_2$ is expected to be semiconductor for optical applications. The calculated band-gaps are listed in Table I. It is clearly that all $AFBiS_2$ are semiconductor with a gap in the range of 0.8-1.0 eV. Note that, DFT-GGA method usually underestimates the band-gap, and there is no experimental data for comparison. To solve this problem, there are a number of methods, such as LDA+U, GW and hybrid functional. The hybrid functional HSE[17] method are extensively used because its predicting band-gap value is very close to the experimental one. Here, the HSE band-gap of $SrFBiS_2$ we calculated is 1.35 eV, which should be closed to its true value. This band-gap value is suitable for solar-cell candidate. Hence, $AFBiS_2$ may be also good photocatalysts for water splitting, degradation and catalysis.

Now, we turn to the birefringent properties that we are interested in. The birefringence (see Fig.1) is found to be frequency-dependent as other optical properties, such as absorption. All $AFBiS_2$ have negative BF when the photonic energy is less than 2 eV. The illustration gives the absolute value of the BF at the energy of zero, i.e., static BF (also listed in Table I). It is evident that all compounds considered here have the larger static birefringence than that of $LaOBiS_2$. Interestingly, $MgFBiS_2$ exhibits the largest BF of 1.66, which is 7.6 times larger than that of $YVO_4$, the most extensively used birefringence laser crystal. The BF value of $MgFBiS_2$ is 66% larger than that of $LaOBiS_2$, which is is 4.5 times larger than $YVO_4$ and is reported to have the largest inherent birefringence among inorganic compounds.[14]

From Mg to Ba, the BFs of $AFBiS_2$ show a very different trend as the photonic energy increasing. It is can be divided to five regions: [0, 2.0], [2.0, 2.9], [2.9, 5.3], [5.3, 9.4] and [9.4, 16]. First region, the range of [0, 2.0] eV, the BF of $AFBiS_2$ decreases as increasing the ionic radii of A-ions, i.e., there is about 26% reduction when A-ion changed from Mg to Ba. For the second region, the range of [2.0, 2.9] eV, the BF remains unchanged for all $AFBiS_2$, indicating they have the same origin which has no connection with AF layers. In fact, it results from the interband transitions between Bi and S ions, as shown as PDOS (Figure 4). The third, the range of [2.9, 5.3] eV, the BF has an opposite trend compared to that in first region. The fourth region has a change of BF from the positive to negative. As for the fifth region, the trend is complex as A-ions changed. When energy is larger than 13 eV, the BF is less than 0.2.

To reveal the origin of giant birefringence (GBF) in $AFBiS_2$, we present the refractive index along [100] and [001] directions, respectively (Fig.2, we make as $n_{100}$ and $n_{001}$, respectively). Note that, the values of $n_{100}$ (3.7-4.3) are evidently larger than the corresponding values (1.9-2.2) of $YVO_4$. This means $AFBiS_2$ have a significant advantage in practical applications. The illustrations show the influence of A-ion radii on the refractive index at the photonic energy of zero. It is evident that $n_{100}$ is linearly reduced as ionic-radii increasing while $n_{001}$ is not (see Fig. 2). Quantitatively, $n_{001}$ is changed only about 0.1, while $n_{100}$ is large up to 0.6. Note that, the value of $n_{001}$ is nearly unchanged for Ca/Sr/Ba ions. The distinguished different influence of A-ions on $n_{100}$ and $n_{001}$ make the BF of $AFBiS_2$ be mainly determined by the change of $n_{100}$. We can conclude that A-ions or AF layers play an important role in the BF of $AFBiS_2$.

To insight the role of A-ions in $AFBiS_2$, we now examine their electronic structures. Figure 3 presents the total density of state (TDOS). It is clear that TDOS has a significant change when energy is lower than -13 eV than that in the range of [-13, 9] eV. The reason may be found in their partial density of state (PDOS). Figure 4a shows that Mg-p orbital sites at -43 eV while Mg-s orbital less than -50 eV. The others A-s orbitals locate at -40, -33 and -25 eV for Ca, Sr and Ba, respectively. Meanwhile, A-p orbital of Ca, Sr and Ba is -21, -15 and -11 eV, respectively. As guided by the dot-lines, the sp-orbitals of A-ions are linearly moved up to Fermi level. This is attributed to the different sp-orbitals of A-ions. Because the F/S2 ions are neighbored with A-ions (see Figure 5), the PDOS of F and S ions are directly affected by A-ions, resulting the F/S-sp orbitals are also upward to Fermi level. The PDOS of Bi ions is affect by the Bi-S1 bonding. The covalent bonding also makes Bi (S)-s,p orbitals move up. In one word, A-ions can turn both A-F and Bi-S bonding through the orbital-hybridizations. Similar phenomenon have been report in $BiMO_3$ ferroelectrics[18] and B-site complex $BiMM'O_3$ compound.[19] Furthermore, the Mulliken charges (Table II) also confirm that A-ions do have a significant influence on other ions.

Electron localization function (ELF) measures the extent of spatial localization of the electrons and provides a method for the mapping of electron pair. The ELF of $AFBiS_2$ is depicted in Fig.5. The anion F and A-cations (Mg, Ca and Ba) have no visible electrons. Here Sr ion is one exception. As expected, the lone pair on $Bi^{3+}$ ions is evident for all $AFBiS_2$. $Bi^{3+}$ lone pair increases in shape, as A-ions changed from Mg to Ca, while remain unchanged from Ca to Sr and then Ba. This is consistent with the trend of $n_{001}$ and may explain the very slight refractive index change for Ca, Sr and Ba (see Figure 2b). The S1 ions in Bi-S plane have also pseudo lone-pair electron in $MgFBiS_2$, while along z axis another pseudo lone-pair appears in other three $AFBiS_2$. The two symmetric pseudo S1 lone-pairs are observable and should result in the reduction of anisotropy of electronic structure. The S2 electrons (down drawing of Fig.5) increase from A-ions Mg changed to Ca, Ba (Sr, exception). This makes the anisotropy of ELF reduce, and should the reason of the reduction of $n_{100}$ (Figure 2a).

Now, we discuss the origin of GBF in detail. Table I indicates that the birefringence of $AFBiS_2$ is reduced when the ion-radii increases as well as the electronegativity reduces. This may be explained by the electronegativity different between A-F/O ions ( EE). The larger EE means there is the larger covalent bonding of A-F. Note that, A-F bonding can turn the Bi-S bonding (Fig.4) as well as Bi/S lone-pair (Fig.5). Therefore, A-ions have an important role in the electronic structure of $AFBiS_2$, in turn in both $n_{100}$ and $n_{001}$. The very different influence on $n_{100}$ and $n_{001}$ result in the large anisotropy, i.e., GBF. Table I shows that Ca and La have nearly the same ionic-radii, while $CaFBiS_2$ has 36% larger BF than $LaOBiS_2$. This means that CaF is more effective than LaO to obtain a large BF. On the other hand, this indicates that A-F/La-O bonding has a significant influence on Bi-S bonding as well as Bi/S lone-pair. Furthermore, $AFBiS_2$ also needs good mechanical properties as a good candidate of birefringent crystal. Because A-ions can effectively turn the electronic structures as demonstrated above, hence, the mechanical properties are proposed to have good tunability. Further experimental exploring on the single-crystal growth and their mechanical character are highly expected.

In summery, we have performed DFT calculations on $AFBiS_2$, which are semiconductor and have superior birefringence than that of $LaOBiS_2$. The DOS and Mulliken charges as well as ELF demonstrate that A-ions play an important role in turning the structural, electric properties and refractive index. A-ions have a very significantly different influence on the refractive index along [100] and [001] directions, resulting in giant birefringence. Finally, $AFBiS_2$ are suggested to have promising potential applications in the fields of laser devices, optical component, solar-cell and photocatalyst.

References:

[1] http://en.wikipedia.org/wiki/Birefringence.

[2] E.G. Villora, K. Shimamura, K. Sumiya, and H. Ishibashi, Opt. Express 17, 12362 (2009).

[3] X. Chen, M.J. Li, N. Venkataraman, M.T. Gallagher, W.A. Wood, A.M. Crowley, J.P. Carberry, L.A. Zenteno, and K.W. Koch, Opt. Express 12, 3888 (2004).

[4] G. Labroille, O. Pinel, N. Treps, and M. Joffre, Opt. Express 21, 21889 (2013).

[5] M.F. Weber, C.A. Stover, L.R. Gilbert, T.J. Nevitt, and A.J. Ouderkirk, Science 287, 2451 (2000).

[6] Y. Yasuno, S. Makita, Y. Sutoh, M. Itoh, and T. Yatagai, Opt. Lett. 27, 1803 (2002).

[7] J.F. deBoer, T.E. Milner, M.J.C. vanGemert, and J.S. Nelson, Opt. Lett. 22, 934 (1997).

[8] C.-Y. Chen, J.-H. Huang, K.-Y. Lai, Y.-J. Jen, C.-P. Liu, and J.-H. He, Opt. Express 20, 2015 (2012).

[9] O.L. Muskens, M.T. Borgstrom, E.P.A.M. Bakkers, and J.G. Rivas, Appl. Phys. Lett. 89, 233117 (2006).

[10] N. Kunzner, D. Kovalev, J. Diener, E. Gross, V.Y. Timoshenko, G. Polisski, F. Koch, and M. Fujii, Opt. Lett. 26, 1265 (2001).

[11] G. Beydaghyan, K. Kaminska, T. Brown, and K. Robbie, Appl. Optics. 43, 5343 (2004).

[12] K. Robbie and M.J. Brett, J. Vac. Sci. Technol. A 15, 1460 (1997).

[13] Y. Terada, K. Shimamura, and T. Fukuda, J. Alloy. Compd. 275, 697 (1998).

[14] H. Wang, arXiv: 1304.5302 (2013).

[15] K.W. H. Lei, M. Abeykoon, E.S. Bozin, and C. Petrovic, arXiv:1208.3189 (2012).

[16] V.P.S. Awana, A. Kumar, R. Jha, S. Kumar Singh, A. Pal, Shruti, J. Saha, and S. Patnaik, Solid State Commun. 157, 21 (2013).

[17] J. Heyd, G.E. Scuseria, and M. Ernzerhof, J. Chem. Phys. 118, 8207 (2003).

[18] H. Wang, B. Wang, Q.K. Li, Z.Y. Zhu, R. Wang, and C.H. Woo, Phys. Rev. B 75, 245209 (2007).

[19] H. Wang, H.T. Huang, W. Lu, H.L.W. Chan, B. Wang, and C.H. Woo, J. Appl. Phys. 105, 053713 (2009).

[20] R.D. Shannon, Acta Cryst. A32, 751 (1976).

Table I. Lattice parameters (a, c in Å), band-gap (G, eV), static birefringence (sBF) at 0 eV, A-ionic radii (IR, Å), the electronegativity of A-ion (EE) and the electronegativitys difference between A-ion and F/O ions (ΔEE=EE$_{A-ion}$-EE$_{F/O-ion}$) and of AFBiS$_2$.

| | a | c | G | sBF | IR[a] | EE | ΔEE |
|---|---|---|---|---|---|---|---|
| MgFBiS$_2$ | 3.816 | 14.047 | 0.79 | 1.66 | 0.72 | 1.31 | 2.7 |
| CaFBiS$_2$ | 3.958 | 14.055 | 0.90 | 1.47 | 1.00 | 1.01 | 3.0 |
| SrFBiS$_2$ | 4.063 | 14.233 | 0.95 | 1.28 | 1.16 | 0.95 | 3.05 |
| | 4.079[b] | 13.814[b] | | | | | |
| BaFBiS$_2$ | 4.182 | 14.579 | 0.94 | 1.10 | 1.35 | 0.89 | 3.1 |
| LaOBiS$_2$ | 4.051 | 14.247 | 0.90 | 1.00 | 1.03 | 1.11 | 2.4 |
| | 4.066[c] | 13.862[c] | | | | | |

[a]Shannon IR, Ref. [20]. [b]Experimental, Ref. [15]. [c]Experimental, Ref. [16].

Table II. Mulliken charges of AFBiS$_2$.

| | A | F/O | Bi | S1 | S2 | n |
|---|---|---|---|---|---|---|
| MgFBiS$_2$ | 1.19 | -0.72 | 0.91 | -0.64 | -0.74 | 1.59 |
| CaFBiS$_2$ | 1.26 | -0.66 | 0.87 | -0.66 | -0.81 | 1.48 |
| SrFBiS$_2$ | 1.18 | -0.63 | 0.88 | -0.67 | -0.76 | 1.34 |
| BaFBiS$_2$ | 1.42 | -0.75 | 0.98 | -0.68 | -0.96 | 1.17 |
| LaOBiS$_2$ | 1.06 | -0.76 | 0.82 | -0.52 | -0.59 | 1.09 |

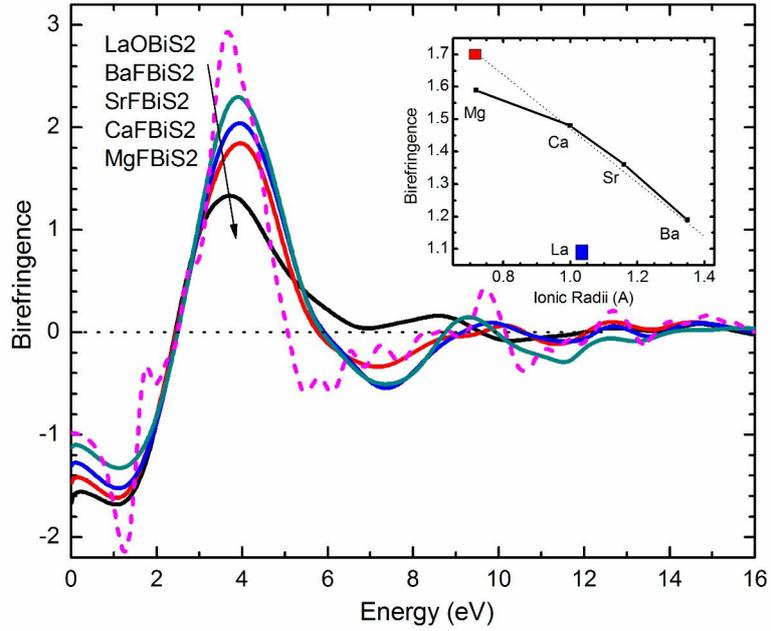

FIG. 1. Birefringence ( n=$n_{zz}$-$n_{xx}$=$n_{001}$-$n_{100}$) of AFBiS$_2$ (Solid-lined) and LaOBiS$_2$ (dot-lined). Insert is the absolute value of the birefringence at 0.5 eV.

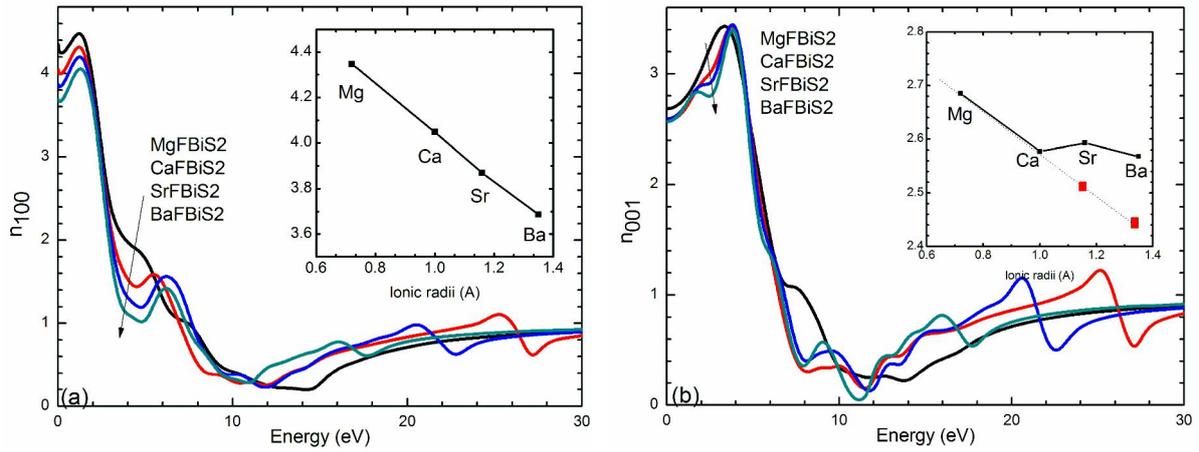

FIG. 2. Refractive index of AFBiS$_2$ along [100] (a) and [001] (b) direction. The values with energy at zero are present in illustration.

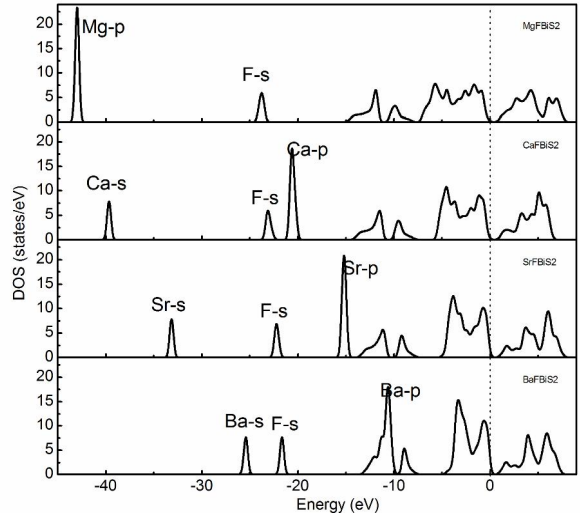

FIG. 3. TDOS of AFBiS$_2$.

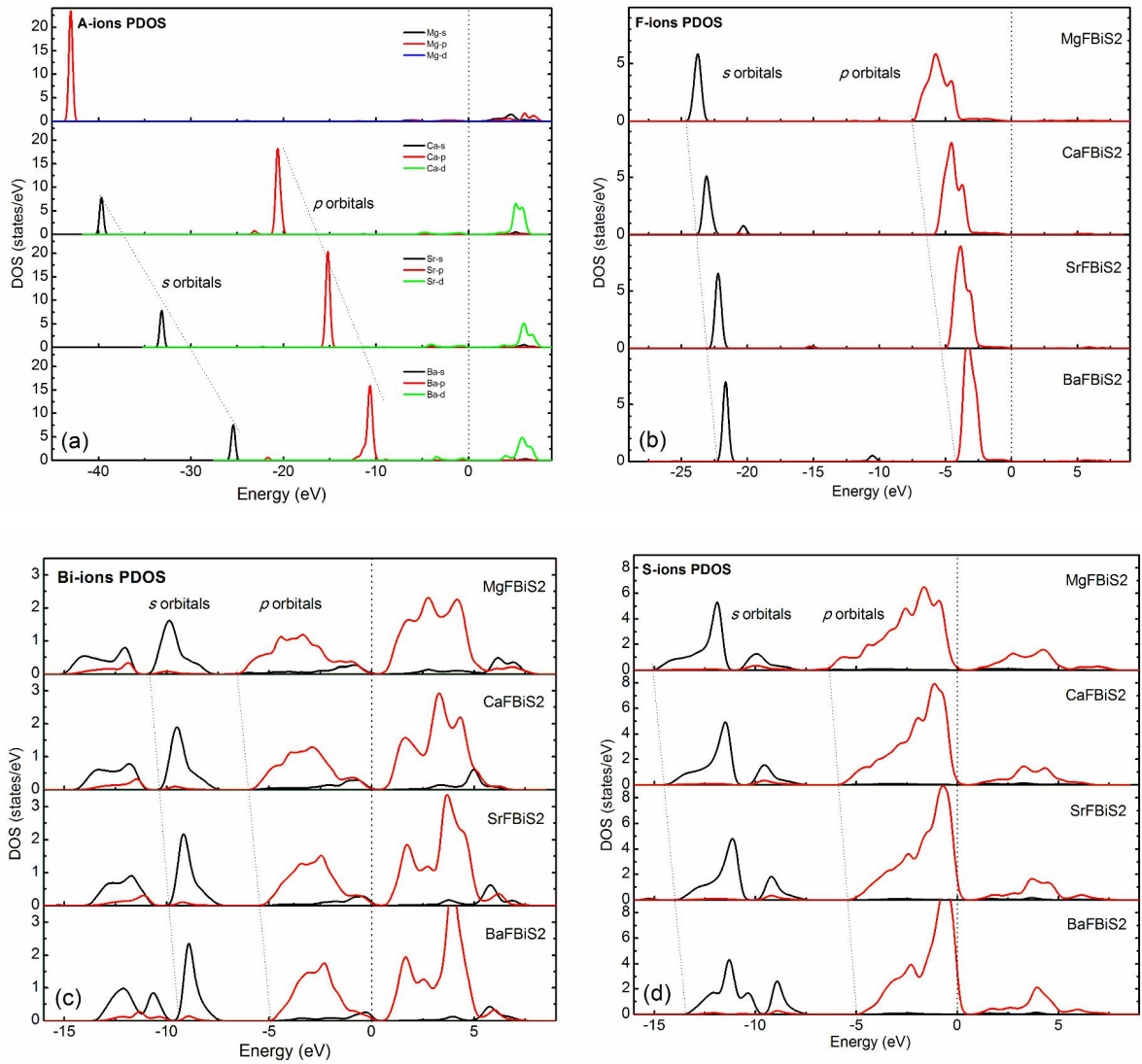

FIG. 4. PDOS of A-ions (a), F (b), Bi (c) and S (d) ions in AFBiS$_2$. Fermi level locates at 0 eV.

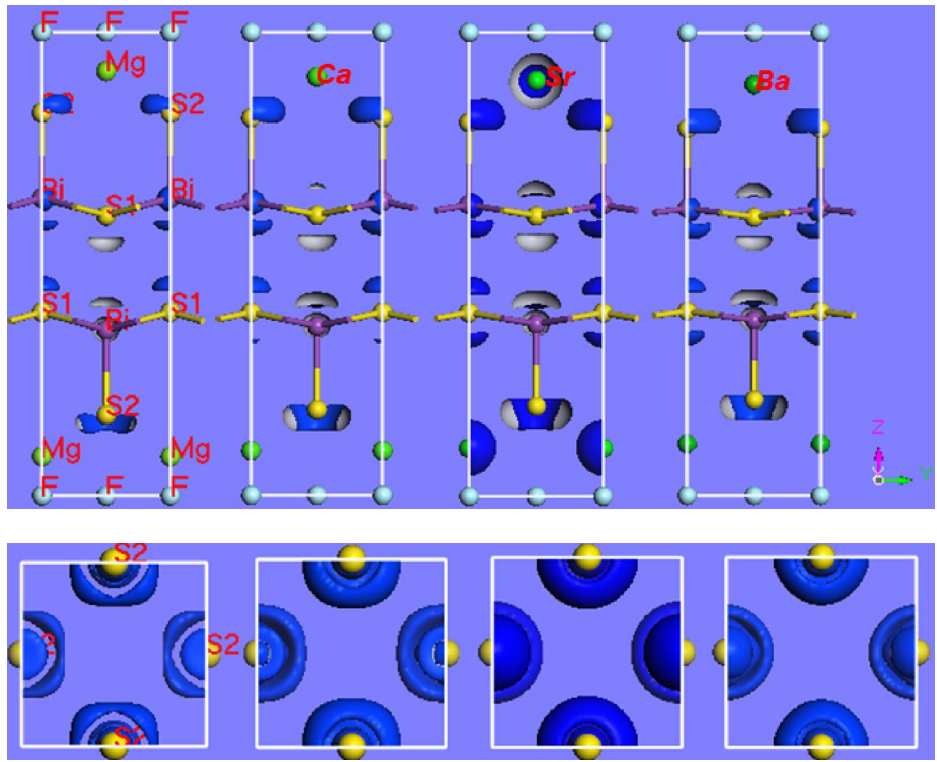

FIG. 5. ELF of AFBiS$_2$: side-view (up) and top-view (down, only show S2 ions).